\documentstyle[12pt,epsfig]{article}
\def\be{\begin{equation}} \def\ee{\end{equation}} 
\def\bea{\begin{eqnarray}} \def\eea{\end{eqnarray}}   
\def\bcc{\begin{center}} \def\ecc{\end{center}}

\def\vf{\varphi} \def\EE{e$^+$e$^-$}  \def\pt{p_{\rm t}}
\def\cl{\centerline} 
 \def\hs{\hskip} \def\vs{\vskip} \def\ni{\noindent}
   
\def\la{\langle} \def\ra{\rangle} 
\begin{document}
\null{}\vskip -2.5cm
\hskip12cm{HZPP-9808}

\hskip12cm{Dec.10, 1998}
\vskip1cm

\begin{center}
{\Large ON A QUALITATIVE DIFFERENCE 
\vskip0.2cm

BETWEEN  THE DYNAMICS OF PARTICLE 
\vskip0.2cm

PRODUCTION IN SOFT AND HARD PROCESSES 
\vskip0.3cm

OF HIGH ENERGY COLLISIONS}

\vskip0.5cm

{\large Liu Feng, \ \ Liu Fuming \ and \ Liu Lianshou}

{\small Institute of Particle Physics, Huazhong Normal University, 
Wuhan 430079 China}
\date{ }
\end{center}

\begin{center}
\begin{minipage}{125mm}
\vskip 0.5in
\begin{center}{\Large ABSTRACT}\end{center}
{\hskip0.6cm  
The qualitative difference between the anomalous scaling properties
of hadronic final states in soft and hard processes of high energy 
collisions is studied in some detail.  It is pointed out that the 
experimental data of \EE collisions at $E_{\rm cm}=$91.2 GeV from 
DELPHI indicate that the dynamical fluctuations in \EE collisions 
are isotropical, in contrast to the anisotropical fluctuations 
oberserved in hadron-hadron collision experiments. This assertion 
is confirmed by the Monte Carlo simulation using the Jetset7.4 
event generator.}
\end{minipage}
\end{center}
\vs0.8cm
{\large PACS number: 13.85 Hd

\ni
Keywords: Multiparticle production, \ Hard and soft processes, \ 

\hskip2.0cm Dynamic fluctuations, \  Self-affine fractal}

\newpage

Quantum Chromodynamics (QCD)~\cite{Altareli} as a candidate of the 
basic theory of strong interaction has been very successful in the 
study of hard processes, such as scaling violation in deep inelastic 
scattering (DIS), hadronic-jet production in \EE annihilation, 
large-$p_{\rm t}$-jet production in hadron-hadron collisions etc.. 
In such processes there exist large scales so that QCD can be solved 
perturbatively due to asymptotic freedom.

On the contrary, in soft-hadronic processes, e.g. hadron-hadron, 
hadron-nucleus and nucleus-nucleus collisions below CERN collider 
energies, there is no large scale present, and the interaction becomes 
so strong that pQCD is no longer applicable.  In these cases, the 
problem is hard to be solved analytically from the first principle and 
phenomenological regularities extracted from experiments have to be 
ultilized to increase our knowledge on the property of the basic 
dynamics in such processes.

In this respect, a comparison of the phenomenology of multiparticle 
final states in soft and hard processes is worthwhile.  Such 
comparison is helpful in getting information about the similarity 
and distinction between the dynamics of particle production in these 
two kinds of processes. In particular, to find out the qualitative 
differences, if any, between them is especially interesting.

In this paper we will show that such a qualitative difference does 
exist in the anomalous scaling property of final-state particle 
distribution. It turns out that the higher-dimensional factorial 
moments (FM) do have anomalous scaling property (obey a power law 
with the diminishing of phase space) when and only when the 
partition of phase space is anisotropic for soft processes while 
isotropic for hard ones. This means that the dynamical fluctuations 
are anisotropic in the former case while isotropic in the latter. 
This qualitative distinction may serve as a useful criterion in the 
study of the basic dynamics of particle production in these two kinds 
of processes. 

Let us first review briefly the status of dynamical-fluctuation study 
in high energy collisions~\cite{Kittel}. The interest in this study 
was first stimulated by the experimental finding in 1983 of 
unexpectedly large local fluctuations (2 times the average) in a high 
multiplicity event from JACEE~\cite{JACEE}.  The same phenomena were 
observed latter also in accelerator experiments, the local 
fluctuations being as large as 60 times the average\cite{NA22spike}. 
Such large fluctuations may not be simply due to statistical reason 
and is an indication of the existence of non-statistical (dynamical) 
fluctuations.

In order to eliminate the statistical fluctuations and study the 
dynamical ones, Bia\l as and Peschanski~\cite{BP} proposed to make use 
of the normalized factorial moments (FM)
\be   %%% (1)
  F_q(M)={\frac {1}{M}}\sum\limits_{m=1}^{M}{{\langle n_m(n_m-1)
     \cdots (n_m-q+1)\rangle }\over {{\langle n_m \rangle}^q}} ,
\ee  
\noindent
where a region $\Delta$ in 1-, 2- or 3-dimensional phase space is 
divided into $M$ cells, $n_m$  is the multiplicity in the $m$th cell, 
and $\langle\cdots\rangle$ denotes vertically averaging over the event 
sample.  They have been able to prove that if dynamical fluctuations 
do exist, the factorial moments will show power-law (anomalous 
scaling) behavior
\be   %%% (2)
   F_q(M) \propto (M)^{\phi_q}\ \  \quad \quad (M\to \infty) \ \ ,
\ee
\noindent
which  can be represented as a straight line in the ln$F_q$ vs. ln$M$ 
plot.  Such kind of power-law scaling is typical for 
fractals~\cite{Mand}.

As known in the fractal theory~\cite{Mand-SA}, when the underlying
space is of higher (2 or 3) dimension, there are qualitatively two
different kinds of fractals --- self-similar and self-affine. The
difference between these two kinds of fractals lies in the scales used 
in different space directions. If one and the same scale is used in
all the directions, i.e. if the power law (2) holds when the space is 
diminished similarly in different directions, the fractal is called
{\em self-similar}. On the contrary, if different scales are used in 
different directions, i.e. if the power law (2) holds when and only 
when the space is diminished with different ratios in different 
directions, the corresponding fractal is called {\em self-affine}. 
An example of the latter is the three dimensional space of landscape, 
where the vertical direction is a special one due to the existence of 
gravity, which causes the vertical variations of landscapes to be 
scaled differently from the horizontal ones.

It turns out that the dynamical fluctuations in the soft-hadronic 
interactions are similar to the 3-D space of landscape in the sense
that the longitudinal direction (the direction of the momenta of
incident hadrons) is privileged so that the higher-dimensional FM
scales when and only when the phase space is partitioned 
anisotropically with different partition numbers in longitudinal and 
transverse directions~\cite{WLPRL}.  In Fig's.1$(a)$ and $(b)$ are 
shown the results from NA22~\cite{NA22SA} and NA27~\cite{NA27SA} 
respectively. 

In order to characterize the different ways of phase space partition, 
a parameter $H$ called Hurst exponent has been used, which is defined 
as
\be  %%% (3)
H_{ab}={\ln M_a\over \ln M_b} ,
\ee
where $M_a$ and $M_b$ are the partition numbers in directions $a$ and 
$b$ respectively. $H=1$ means that the phase space is divided 
similarly in directions $a$ and $b$, while $H\neq 1$ characterizes the 
anisotropical way of phase space partition, cf. Fig.2.

It can be seen from Fig's.1$(a)$ and $(b)$ that the log-log plots of 
the second order FM ($F_2$) versus the partition number ($M$) in 
hadron-hadron collisions, instead of being straight, are bending
upwards when the phase space is divided similarly in different 
directions ($H=1$). On the other hand, if the phase space is divided
anisotropically with some particular values of $H$ ---
$H_{\parallel\perp}=0.475, H_{p_{\rm t}\varphi}=1$ for NA22, 
$H_{\eta\varphi}=0.74$ for NA27 --- the log-log plots become straight
within the experimental error. These results show that the dynamical
fluctuations in soft-hadronic interactions are anisotropical and the
corresponding fractal is self-affine.

However, the published results of hard collisions, e.g. the results of
\EE annihilation at $E_{\rm cm}=91.2$GeV from DELPHI~\cite{delphi}, 
showed a sharply different situation. In this case, the 3-D ln$F_2$ 
vs. ln$M$ plot under isotropical phase space partition ($H=1$) follows 
a straight line reasonably well, cf. Fig.1$(c)$, provided the first 
point is omitted to reduce the influence of momentum 
conservation~\cite{MMCN}. This remarkable fact indicates that 
dynamical fluctuations in soft and hard processes are qualitatively 
different, being anisotropic in the former case while isotropic in 
the latter.

In order to confirm this observation, we have done a Monte Carlo 
simulation with Lund JETSET7.4 event generator. In total 1,000,000 
events are produced for \EE collisions at $E_{\rm cm}=91.2$GeV. 
The FM analysis is done both in the laboratory system and in the 
thrust system. In order to reduce the trivial effect of the non-flat
of average distributions the cummulant variables have been 
used~\cite{CUMVAR}.

In Fig.3 and Table~I are shown the results for lab. system. In this 
system the $z$ axis is chosen to be along the direction of motion of 
the incident particles (e$^+$ and e$^-$). Since these two particles 
have been annihilated into a single virtual particle (Z$^0$ or 
$\gamma^*$), their direction of motion is not privileged in the final 
state hadronic system any more. Therefore, it is meaningless to 
distinguish ``longitudinal'' and ``transverse'' directions in this 
coordinate system and so the Cartesian momenta $p_x, p_y, p_z$, 
instead of the ``longitudinal Lorentz invariant'' variables $y, 
p_{\rm t}, \vf$, are used.

\begin{center}
{\bf Table~I} \ \ The fitting parameters of 1-D FM in the lab. system
\vskip0.8cm

\small
\begin{tabular}{|c|c|c|c|c|}\hline
\mbox{ Varibles} &$ A $ & $ B $ &$ \gamma$ & \mbox{Omitting point(s)} 
\\  \hline
 $p_x$ & 1.217 $\pm$ 0.001 & 0.400 $\pm$ 0.004 & 1.278 $\pm$ 0.010 & 
1 \\  
\hline
 $p_y$ & 1.218 $\pm$ 0.001 & 0.399 $\pm$ 0.004 & 1.267 $\pm$ 0.009 & 
1 \\  
\hline
 $p_z$ & 1.214 $\pm$ 0.001 & 0.409 $\pm$ 0.004 & 1.368 $\pm$ 0.010 & 
1 \\  
\hline
\end{tabular}
\bigskip
\end{center}
\vskip1cm

It can be seen from Fig.3$a$ that the log-log plot of 3-D $F_2$ versus
$M$ fits a straight line well. The value of Hurst exponents can be 
obtained through fitting the three 1-D plots, cf. Fig.3$b$, to the 
formulae~\cite{ZGKX}
\be   %%% (4)
 F_2^{(a)}(M_a) = A_a-B_a M_a^{-\gamma_a}, \ \ a=p_x,p_y,p_z
\ee
as
\be   %%% (5)
 H_{ab}\equiv {\ln\lambda_a \over \ln\lambda_b} = {1+\gamma_b\over
  1+\gamma_a}. 
\ee
The values of fitting parameters are listed in Table~I and the 
resulting Hurst exponents:
$$H_{p_xp_y}={1+\gamma_{p_y}\over 1+\gamma_{p_x}}=0.995 \pm 0.008, $$
$$H_{p_zp_y}={1+\gamma_{p_y} \over 1+\gamma_{p_z}} =0.957 \pm 0.008,$$ 
$$H_{p_zp_x}={1+\gamma_{p_x} \over 1+\gamma_{p_z}} =0.962 \pm 0.008 $$ 
are nearly equal to unity. All of these show clearly that the 
dynamical fluctuations are isotropical and the corresponding fractal 
is self-similar rather than self-affine.

As is well known, in high energy \EE collisions hadronic jets
are produced. The reason is that the virtual photon or Z$^0$ formed in
\EE annihilation first produce a pair of fast moving quark-antiquark,
and then the latter radiating gluons turned finally into hadrons.
The direction of motion of the quark-antiquark pair is privileged in
the final state hadronic system and can be chosen as the 
``longitudinal'' direction. The longitudinal rapidity 
\be   %%% (6)
   y=\ln {1 \over 2}{E+p_\parallel \over E-p_\parallel} 
\ee
can thus be defined, where $p_\parallel$ is the momentum along the 
``longitudinal'' direction. 

Experimentally, the direction of the quark-antiquark pair can be 
extracted from the momenta of the final-state particles as the 
direction determined by~\cite{bra64}
\be   %%% (7)
  T=\max {\sum_i |p_{\parallel i}| \over \sum_i |\vec p_i|}  .
\ee
This direction is called {\em thrust axis}, and can be taken as the 
``longitudinal'' 
direction. Using the same formula in the plan 
perpandicular to $T$, the {\em major axis} $T_2$~\cite{mar79} is 
obtained. The direction perpendicular to both $T$ and $T_2$ is called 
the {\em minor axis} $T_3$.

The {\em thrust coordinate system} is defined as a Cartesian 
coordinate system, using the three axes $T_3$, $T_2$ and $T$ as 
$x$, $y$ and $z$ axes respectively. The longitudinally Lorentz 
invariant variables $y$, $p_{\rm t}$ and $\vf$ are then obtained with 
the rapidity $y$ defined according to Eq.(5). The azimuthal angle 
$\vf$ is defined relative to the $x$ ($T_3$) axis.

\begin{center}
{\bf Table~II} \ \ The fitting parameters of 1-D FM in the thrust 
system

\bigskip
\small
\begin{tabular}{|c|c|c|c|c|}\hline
\mbox{Varibles} &$ A $ & $ B $ &$ \gamma$ & \mbox{Omitting point(s)} 
\\  \hline
  $y$ & 1.468 $\pm$  0.001 & 0.751 $\pm$ 0.003 & 0.887 $\pm$ 0.004 & 
1 \\  
\hline
 $\pt$ & 1.157 $\pm$  0.001 & 0.130 $\pm$ 0.002 & 0.825 $\pm$ 0.017 & 
1 \\  
\hline
$\vf$ & 1.033 $\pm$  0.001 & 1.077 $\pm$ 0.037 & 2.742 $\pm$ 0.027 & 
3 \\  
\hline
\end{tabular}
\bigskip
\end{center}

The resulting log-log plots are shown in Fig.4 and the fitting 
parameters are listed in Table~II. It turns out that, inspite of the 
3-D log-log plots being still straight, the three 1-D plot are very 
different, so that the three sets of fitting parameters differ 
sharply. In particular, the value of $\gamma_\vf$  is more than 3 
times bigger than $\gamma_y$ and $\gamma_{p_{\rm t}}$, which would 
mean $H_{y\vf}, H_{p_{\rm t}\vf} \gg 1$, cf. Eq.(5).  However, this 
is not really the case.

Let us notice that there are 3 parameters in the projection formula 
Eq.(4).  In getting the value of $H$, only the fitting parameters 
$\gamma$ are used, cf. Eq.(5), but this does not mean that the values 
of the other two parameters $A$ and $B$ are unimportant. The 
projection formulae for these two parameters are~\cite{OCHSproj}:
\be   %%% (8)
A={\lambda -1\over \lambda - C^{(2)} }, \qquad
B={C^{(2)} -1\over \lambda - C^{(2)} }, \qquad
\ee
where $C^{(2)}={\la \omega^2\ra / \la\omega\ra^2}$ is the normalised 
second order moment of the probability $\omega$ of elementary
space partition; $\lambda$ is the elementary partition number. It is 
evident that the parameters $A$ and $B$ satisfy a relation:
\be  %%% (9)
A-B=1.
\ee
This relation should be approximately valid for any physically 
meaningful set of parameters. In Fig.5 are shown the difference
$A-B$ for all the presently available cases. It can be seen that,
in all the cases except for variable $\vf$ in the thrust system 
of Jetset, relation (9) holds approximately, while for the latter
case, this relation is sharply violated. This means that the set
of fitting values listed in Table~II is physically meaningless, 
and therefore it is impossible to draw any conclusion about the value 
of Hurst exponent from it.

The reason for the exceptional nature of the dynamical fluctuations
for variable $\vf$ in the thrust system is understandable. It is 
because the relative point for counting the value of $\vf$, i.e. the
$x$ axis, in the thrust system is fixed at the third thrust axis
$T_3$. As a crude estimation, the second thrust axis $T_2$ is
approximately the direction of the first hard gluon
emitted by the quark or antiquark. To count $\vf$ from $T_3$ 
means that the azimuthal angle of first hard gluon emission is fixed
to 90 degree. Thus the fluctuations of the direction of first hard gluon 
emission are largely reduced.

In order to study the real 3-D fluctuations, it is necessary to loosen
the correlation between the direction of first hard gluon emission and
the $x$ axis.
For this purpose, putting the $z$ axis still on the main thrust axis, 
we turn the coordinate system arround it and let the new $x$ axis 
lies on the $z_0$-$z$ plan, where $x_0,y_0,z_0$ denote the axes of 
the lab system and $x,y,z$ those of the turned system, as shown in Fig.6. 

The log-log plots in the turned system are shown in Fig.7 and the
fitting parameters are listed in Table~III. 
The 3-D plot is still straight. The three 1-D plots become similar
and the differences between the fitting parameters $A$ and $B$ for 
these three plots, especially that for the variable $\vf$, become near 
to unity, cf. Fig.5.  The Hurst exponents calculated from the fitting 
parameters:
$$H_{yp_{\rm t}}={1+\gamma_{p_{\rm t}} \over 1+\gamma_{y}} =0.95 \pm 
0.02,  $$
$$H_{yp_\vf}={1+\gamma_{\vf} \over 1+\gamma_{y}} =1.11 \pm 0.02, $$ 
$$H_{p_{\rm t}\vf}={1+\gamma_{\vf} \over 1+\gamma_{p_{\rm t}}} =1.18 
\pm 0.03 $$ 
are approximately equal to unity. This confirms further that the 
dynamical fluctuations are isotropical in \EE collisions and the 
corresponding fractal is self-similar.

\begin{center}
{\bf Table~III} \ \ The fitting parameters of 1-D FM in the turned 
system

\bigskip
\small
\begin{tabular}{|c|c|c|c|c|}\hline
\mbox{Varibles} &$ A $ & $ B $ &$ \gamma$ & \mbox{Omitting point(s)} 
\\  \hline
   $y$ & 1.469  $\pm$       0.001    &  
  0.755    $\pm$     0.002   &  
  0.890     $\pm$    0.004  & 1 \\ 
\hline
  $\pt$ & 1.160      $\pm$   0.001  &  
  0.131     $\pm$    0.002 &  
  0.785    $\pm$     0.014    & 1 \\ 
\hline
 $\vf$ & 1.175     $\pm$    0.001 &  
  0.492     $\pm$    0.008  &  
  1.102    $\pm$     0.010     &  3 \\ 
\hline
\end{tabular}
\bigskip
\end{center}

In conclusion, in this paper a qualitative difference between the 
dynamics of particle production in soft and hard processes of high 
energy collisions is analysed in some detail. 
It is pointed out that the experimental data of \EE collisions at 
$E_{\rm cm}=$91.2 GeV from DELPHI indicate that the dynamical 
fluctuations in the hadronic final states of this experiment are 
isotropical, in contrast to the ansotropical fluictuations oberserved 
in hadron-hadron collision experiments (NA22 and NA27). This 
observation is confirmed using the Jetset7.4 event generator. The 
3-D factorial moments (FM) in the Monte Carlo sample with 1000000 
events at $E_{\rm cm}=$91.2 GeV shows good scaling when the phase 
space is divided isotropically. The three 1-D FM's in the lab frame 
have similar shapes. The corresponding Hurst exponents are nearly 
equal to unity. In the thrust system, after turning the transverse 
axes to eliminate the correlation with the direction of first hard 
gluon emission, the 1-D FM's become similar too, and the Hurst 
exponents are approximately equal to unity.

Let us discuss briefly why the multiparticle final states in \EE 
collisions are isotropic (self-similar) instead of anisotropic 
(self-affine) fractal.

It has long been recognized~\cite{Veneziano} that QCD branching has 
fractal structure. Being a hard process with large $Q^2$, the 
``branches'' are developed isotropically in 3-D phase space without 
a limited $p_{\rm t}$~\cite{Veneziano}. The resulting parton 
distribution is naturally a self-similar (isotropic) fractal. What 
are observed in experiments are, of course, not the partons 
themselves, but the final state hadrons produced from them. However, 
the hadronization of each parton is a soft process with limited 
$p_{\rm t}$ perpendicular to the jet axis. These axes are distributed 
in 3-D phase space as an isotropic fractal, without privileged 
direction. Therefore, the anisotropic effects of the hadronization of 
different partons cancell each other, and the isotropic (self-similar) 
fractal of parton-distribution will be largely survived after 
hadronization. This is why an approximately self-similar fractal can 
be observed in \EE collisions. 

The qualitatively different fractal properties between the hadronic 
final states of hard and soft collisions (self-similar and 
self-affine) may serve as a useful criterion in the study of the 
basic dynamics of particle production in these two kinds of processes. 

\vskip0.5cm

\ni Acknowledgement

This work is supported in part by the NSFC. The authors are grateful 
to W. Kittel and Wu Yuanfang for helpful discussion. 

\newpage

%%%%%%%%%%%%%%%%%%%%%%%%%%%%%%%%%%%%%%%%%%%%%%%%%%%%%%%%%%%%%%%%%%%%%%
\def\Journal#1#2#3#4{{#1} {\bf #2}, #3 (#4)}
\def\NCA{\em Nuovo Cimento} \def\NIM{\em Nucl. Instrum. Methods}
\def\NIMA{{\em Nucl. Instrum. Methods} {\bf A}} 
\def\NPB{{\em Nucl. Phys.} {\bf B}}
\def\PLB{{\em Phys. Lett.} {\bf B}} \def\PRL{\em Phys. Rev. Lett.}
\def\PRD{{\em Phys. Rev.} {\bf D}} \def\ZPC{{\em Z. Phys.} {\bf C}}
%%%%%%%%%%%%%%%%%%%%%%%%%%%%%%%%%%%%%%%%%%%%%%%%%%%%%%%%%%%%%%%%%%%%%%

\newpage

\ni{\Large\bf Figure Captions}
\vskip0.8cm

{{\bf Fig.1} \ Qualitative difference in the scaling property 
of factorial moments in h-h 
\vskip0.2cm

\qquad \ \ \ and e$^+$e$^-$ collisions.
\vskip0.2cm

{{\bf Fig.2} \ A sketch of the different ways of phase space division.

\qquad \ \ \ $(a)$ isotropic; $(b)$ anisotropic, $H<1$; 
$(c)$ anisotropic, $H>1$.}
\vskip0.2cm

{{\bf Fig.3} \ Factorial moments of Jetset Monte Carlo sample in the 
lab frame.}
\vskip0.2cm

{{\bf Fig.4} \ Factorial moments of Jetset Monte Carlo sample in the 
thrust frame}
\vskip0.2cm

{{\bf Fig.5} \ Compilation of the values of $A-B$. }
\vskip0.2cm

{{\bf Fig.6} \ The turned coordinate system.}
\vskip0.2cm

{{\bf Fig.7} \ Factorial moments of Jetset Monte Carlo sample in the 
turned frame.}
\vskip0.2cm

\newpage

\null{}\vskip3cm

\begin{picture} (260,240) 
\put(-140,-230)   %%  \center
{\epsfig{file=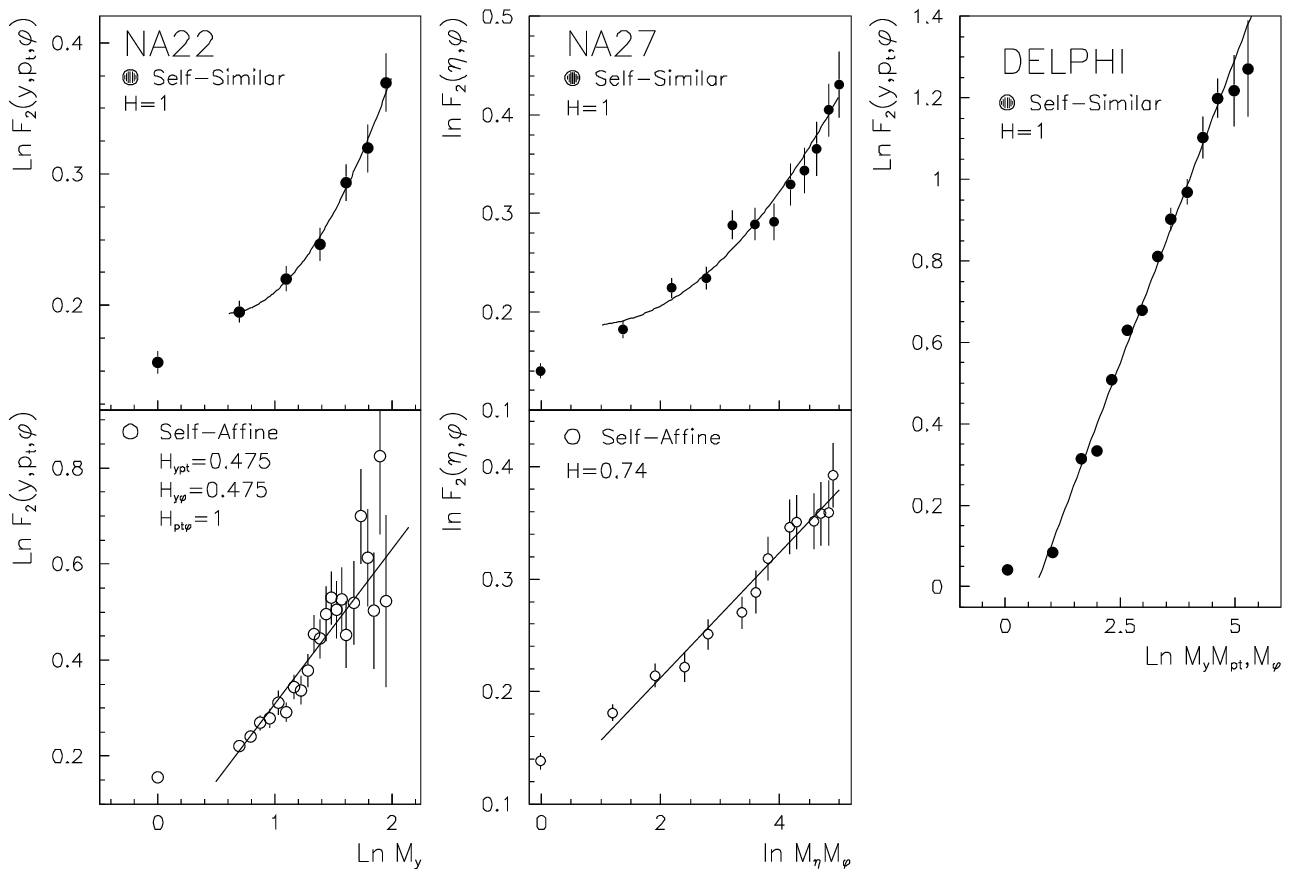,bbllx=0cm,bblly=0cm,
           bburx=8cm,bbury=6cm}}  %%  height=1.5in}
\end{picture}
\vs-4.0cm
\hs1.2cm $(a)$ \hs3.5cm $(b)$ \hs3.8cm $(c)$
\vskip0.5cm
\cl{{\bf Fig.1}}
%%% \cl{{\bf Fig.1} \ Qualitative difference in the scaling property }
%%% \cl{ of factorial moments in h-h and e$^+$e$^-$ collisions}
\vs-13cm
\begin{picture} (260,240) 
\put(-120,-600)   %%  \center
{\epsfig{file=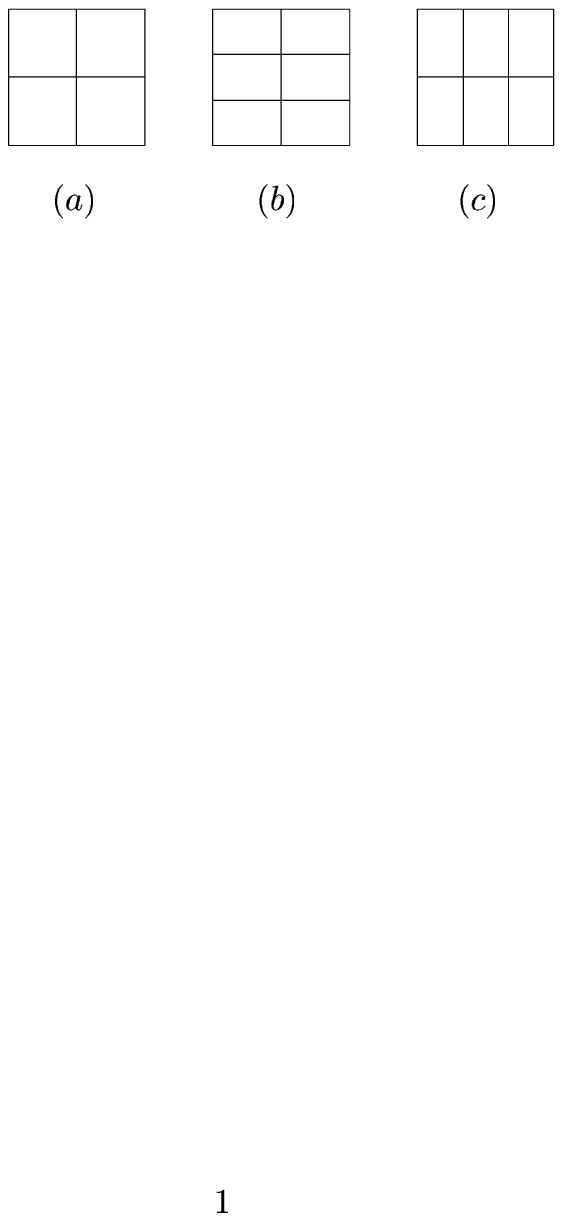,bbllx=0cm,bblly=0cm,
           bburx=8cm,bbury=6cm}}  %%  height=1.5in}
\end{picture}

\vs8.0cm
%%% \hs3.6cm $(a)$ \hs1.5cm $(b)$ \hs1.5cm $(c)$

\cl{ {\bf Fig.2}}
%%% \cl{ {\bf Fig.2} \ Different ways of phase space division}
%%% \cl{ $(a)$ isotropic; $(b)$ anisotropic, $H<1$; 
%%% $(c)$ anisotropic, $H>1$}
\vs-5.0cm

\begin{picture}(260,240)
\put(-145,-580)
{\epsfig{file=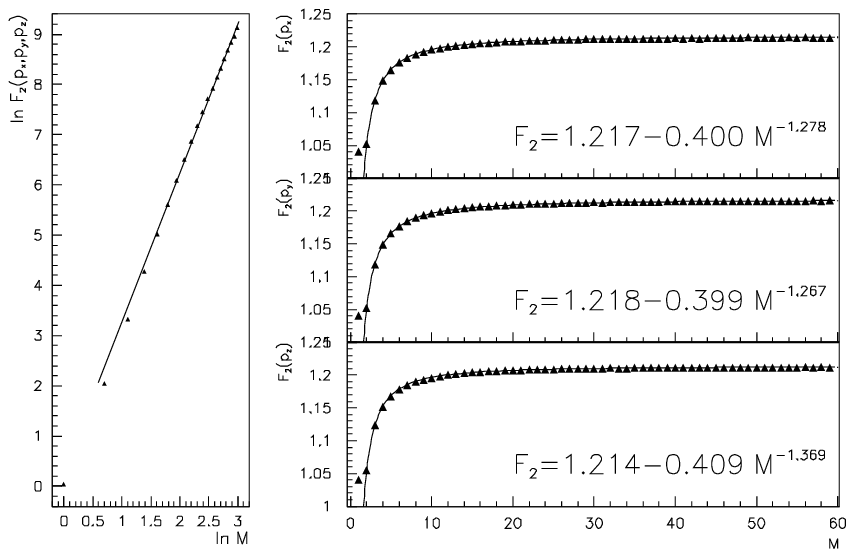,bbllx=0cm,bblly=0cm,
	   bburx=8cm,bbury=6cm}}
\end{picture}

\vs2.5cm
\hs3.2cm $(a)$ \hs4.0cm $(b)$ 

\vs 0.1cm
\cl{ {\bf Fig.3}}
%%% \cl{ {\bf Fig.3} \ Factorial moments of Jetset Monte Carlo sample 
%%% in the lab frame}
\newpage

\null{}\vskip2cm

\begin{picture}(260,240)
\put(-165,-400)    %%   \put(-205,-180)
{\epsfig{file=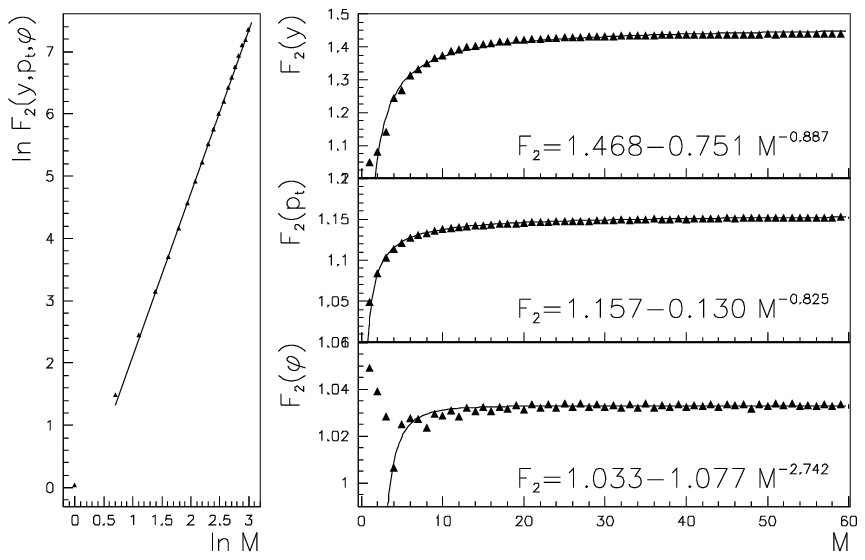,bbllx=0cm,bblly=0cm,
	   bburx=8cm,bbury=6cm}}
\end{picture}

\vs-2.6cm
\cl{{\bf Fig.4}}
%%% \cl{{\bf Fig.4} \ Factorial moments of Jetset Monte Carlo sample 
%%% in the thrust frame}

\begin{picture} (260,240) 
\put(-90,-360)   %%  \put(-140,-260)   
{\epsfig{file=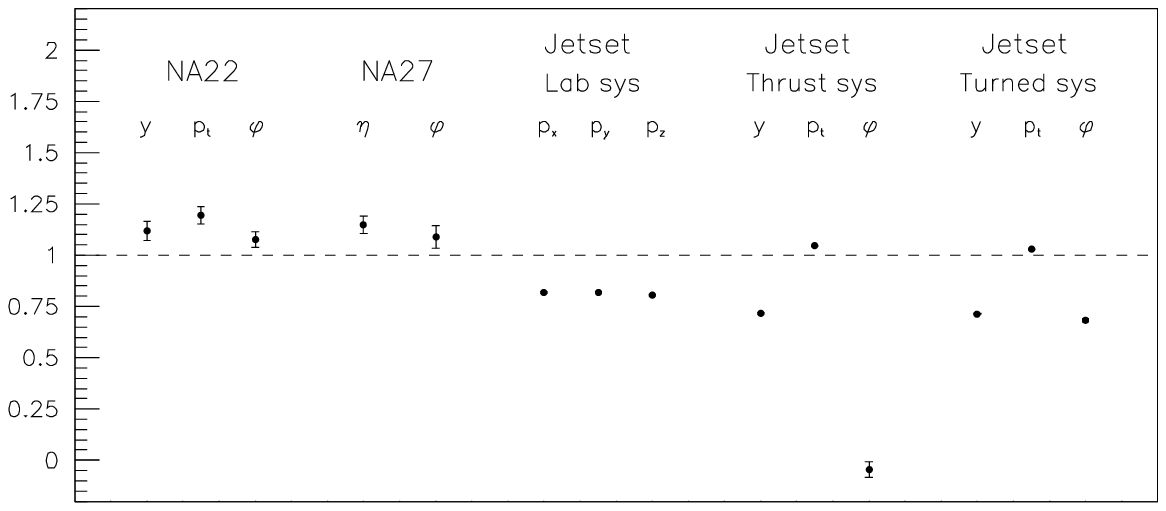,bbllx=0cm,bblly=0cm,
           bburx=8cm,bbury=6cm}}  %%  height=1.5in}
\end{picture}
\vs1.5cm
\cl{{\bf Fig.5}} 
%%% \cl{{\bf Fig.5} \ Compilation of the values of $A-B$ }

\newpage

\begin{picture} (260,240) 
\put(-140,-330)     %%  \put(-140,-330)   
{\epsfig{file=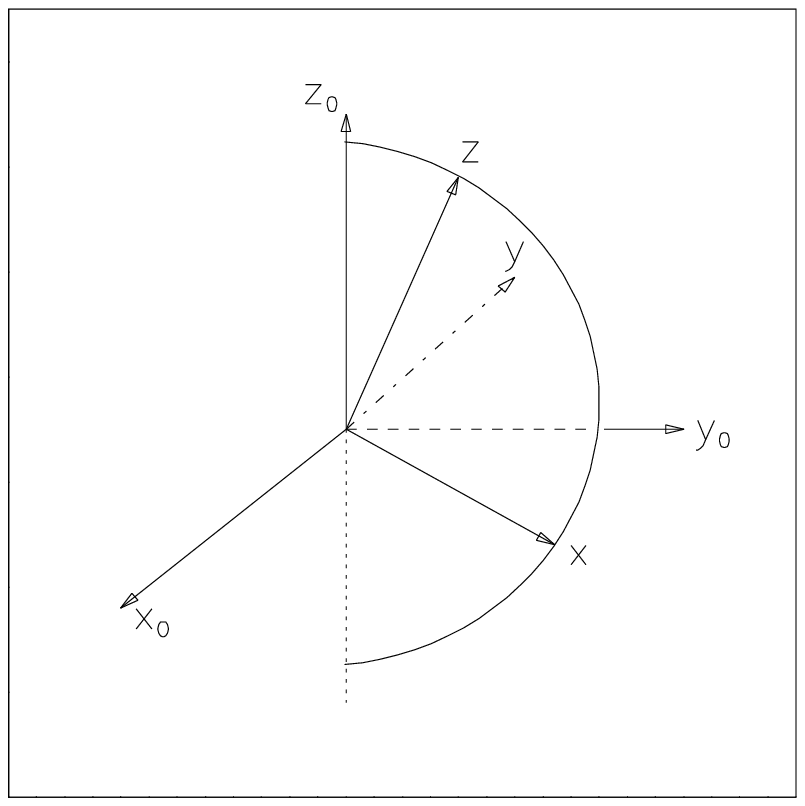,bbllx=0cm,bblly=0cm,
           bburx=8cm,bbury=6cm}}  %%  height=1.5in}
\end{picture}

 \vs2.0cm
\cl{{\bf Fig.6}} 
%%% \cl{{\bf Fig.6} \ The turned coordinate system}

\begin{picture}(260,240)
\put(-150,-500)
{\epsfig{file=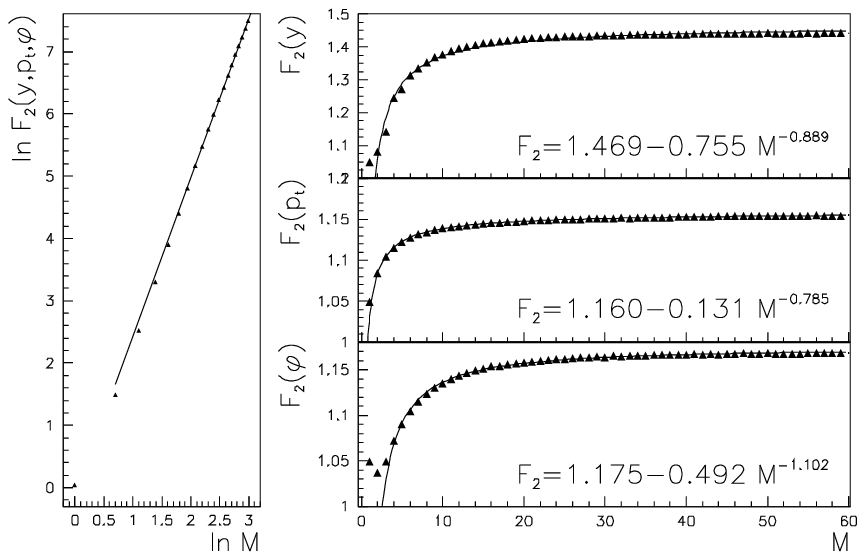,bbllx=0cm,bblly=0cm,
	   bburx=8cm,bbury=6cm}}
\end{picture}

\vs0.5cm
\cl{{\bf Fig.7}} 


\begin{thebibliography}{99}

\bibitem{Altareli} See for example G. Altarelli, {\em The Developement
of Perturbative QCD}, World Scientific, Singapore 1994.

\bibitem{Kittel} See for example the review article: E. A. De Wolf,
I. M. Dremin and W. Kittel, \Journal{\em Phys. Rep.}{270}{1}{1996}.

\bibitem{JACEE} T. H. Burnett et al., \Journal{\PRL}{50}{2062}{1983}.  

\bibitem{NA22spike} M. Adamus et al. (NA22), 
\Journal{\PLB}{185}{200}{1987}.

\bibitem{BP} A. Bia\l as and R. Peschanski, 
\Journal{\NPB}{273}{703}{1986}; \Journal{\em ibid}{308}{857}{1988};

\bibitem{Mand} B. Mandelbrot, {\em The Fractal Geometry of Nature} 
(Freeman, NY, 1982).

\bibitem{Mand-SA} B. Mandelbrot in {\em Dynamics of Fractal Surfaces}, 
ed. E. Family and T. Vicsek (World Scientific, Singapore, 1991).

\bibitem{WLPRL} Wu Yuanfang and Liu Lianshou, \Journal
{\PRL}{21}{3197}{1993}.

\bibitem{NA22SA} N. M. Agabayan et al. (NA22), 
\Journal{\PLB}{382}{305}{1996}; N. M. Agabayan et al. (NA22), 
\Journal{\PLB}{431}{451}{1998};

\bibitem{NA27SA} S. Wang, Z. Wang and C. Wu, 
\Journal{\PLB}{410}{323}{1997}.

\bibitem{delphi} P. Abreu (DELPHI), {\em Nucl. Phys.} {\bf B386} 
471 (1992).

\bibitem{MMCN}
Liu Lianshou, Zhang Yang and Deng Yue, 
\Journal{\ZPC}{73}{535}{1997}.

\bibitem{CUMVAR} W. Ochs, \Journal{\ZPC}{50}{339}{1991}.

\bibitem{ZGKX} Wu Yuanfang and Liu Lianshou, 
\Journal{{\em Science in China} (series A)}{38} {435}{1995}.

\bibitem{bra64} S. Brandt, Ch. Peyrou, R. Sosnowski and A. Wroblewski,
{\em Phys. Lett.} {\bf 12} 57 (1964).

\bibitem{mar79} MARK J Coll. D. P. Barber et al., 
\Journal{PRL}{43}{830}{1979}. 

\bibitem{OCHSproj} W. Ochs, \Journal{\PLB}{347}{101}{1990}.

\bibitem{Veneziano} G. Veneziano, talk given at the {\em 3rd Workshop 
on Current Problems in High Energy Particle Theory}, Florence, 1979.

\end{thebibliography}
\end{document}